\documentclass[
12pt,
preprint,preprintnumbers,nofootinbib,
groupedaddress,superscriptaddress,amsmath,amssymb]{revtex4}
\usepackage{graphicx}
\usepackage{dcolumn}
\usepackage{bm}
\usepackage{amssymb}
\usepackage{amsmath}
\usepackage{epsfig}    
\usepackage{color}
\usepackage{hhline}

\def\be{\begin{equation}}
\def\ee{\end{equation}}
\newcommand{\bea}{\begin{eqnarray}}
\newcommand{\eea}{\end{eqnarray}}
\newcommand{\nn}{\nonumber}

\numberwithin{equation}{section}

\begin{document}
 {\begin{flushright}{KIAS-P19067, APCTP Pre2019 - 026}\end{flushright}}
\title{An Inverse Seesaw model with $A_4$-modular symmetry}
%

\author{Takaaki Nomura}
\email{nomura@kias.re.kr}
\affiliation{School of Physics, KIAS, Seoul 02455, Republic of Korea}

\author{Hiroshi Okada}
\email{hiroshi.okada@apctp.org}
\affiliation{Asia Pacific Center for Theoretical Physics (APCTP) - Headquarters San 31, Hyoja-dong,
Nam-gu, Pohang 790-784, Korea}
\affiliation{Department of Physics, Pohang University of Science and Technology, Pohang 37673, Republic of Korea}

\author{Sudhanwa Patra}
\email{sudhanwa@iitbhilai.ac.in}
\affiliation{Department of Physics, Indian Institute of Technology Bhilai, Raipur-492015, Chhattisgarh, India}


\begin{abstract}
{
We discuss an inverse seesaw model based on right-handed fermion specific $U(1)$ gauge symmetry and $A_4$-modular symmetry. 
These symmetries forbid unnecessary terms and restrict structures of Yukawa interactions which are relevant to inverse seesaw mechanism.
Then we can obtain some predictions in neutrino sector such as Dirac-CP phase and sum of neutrino mass, which are shown by our numerical analysis.
Besides the relation among masses of heavy pseudo-Dirac neutrino can be obtained since it is also restricted by the modular symmetry.
We also discuss implications to lepton flavor violation and collider physics in our model.
}
\end{abstract}
\maketitle
\newpage
\section{Introduction}\label{sec1}


One of the challenging issue in particle physics is the understanding of flavor structure of fermions in the standard model (SM).
In the SM, we do not have any principle to determine the structure and we expect it can be explained in a framework of new physics beyond the SM.
In constructing a new physics model to describe the flavor structure, a new symmetry can play an important role to control the structure of flavors.

Recently the framework of modular flavor symmetries have been proposed by~\cite{Feruglio:2017spp, deAdelhartToorop:2011re}
to realize more predictable flavor structures. 
In this framework, a coupling can be transformed under a non-trivial representation of a non-Abelian discrete group and many scalar fields such as flavons are not necessary to realize flavor structure.  
Then some typical groups are found to be available in basis of  the $A_4$ modular group \cite{Feruglio:2017spp, Criado:2018thu, Kobayashi:2018scp, Okada:2018yrn, Nomura:2019jxj, Okada:2019uoy, deAnda:2018ecu, Novichkov:2018yse, Nomura:2019yft, Okada:2019mjf,Ding:2019zxk,Nomura:2019lnr,Kobayashi:2019xvz,Asaka:2019vev,Zhang:2019ngf, Gui-JunDing:2019wap}, $S_3$ \cite{Kobayashi:2018vbk, Kobayashi:2018wkl, Kobayashi:2019rzp, Okada:2019xqk}, $S_4$ \cite{Penedo:2018nmg, Novichkov:2018ovf, Kobayashi:2019mna,King:2019vhv,Okada:2019lzv,Criado:2019tzk,Wang:2019ovr}, $A_5$ \cite{Novichkov:2018nkm, Ding:2019xna,Criado:2019tzk}, larger groups~\cite{Baur:2019kwi}, multiple modular symmetries~\cite{deMedeirosVarzielas:2019cyj}, and double covering of $A_4$~\cite{Liu:2019khw} in which  masses, mixing, and CP phases for quark and/or lepton are predicted.~\footnote{Some reviews are useful to understand the non-Abelian group and its applications to flavor structure~\cite{Altarelli:2010gt, Ishimori:2010au, Ishimori:2012zz, Hernandez:2012ra, King:2013eh, King:2014nza, King:2017guk, Petcov:2017ggy}.}
A possible correction from K\"ahler potential is also discussed in Ref.~\cite{Chen:2019ewa}.
Furthermore, a systematic approach to understand the origin of CP transformations has been recently discussed in ref.~\cite{Baur:2019iai}, 
and CP violation in models with modular symmetry is discussed in Ref.~\cite{Kobayashi:2019uyt}.
In particular, it is interesting to apply a modular symmetry in constructing a new physics model for neutrino mass generation since 
we would obtain prediction for signals of new physics correlated with observables in neutrino sector.

In this paper, we apply modular $A_4$ symmetry to the inverse seesaw mechanism which is realized introducing new local Abelian gauge symmetry, denoted as $U(1)_R$, which is right-handed fermion specific~\cite{Nomura:2018mwr}.
 The inverse seesaw mechanism requires a left-handed neutral fermions $S_L$ in addition to the right-handed ones $N_R$, and provides us more
 complicated neutrino mass matrix which can make mass hierarchies softer than the other models such as canonical seesaw~\cite{Seesaw1, Seesaw2, Seesaw3, Seesaw4} and provide rich phenomenologies such as unitarity constraints~\cite{Mohapatra:1986bd, Wyler:1982dd}. 
The $U(1)_R$ symmetry requires three SM singlet fermions with non-zero $U(1)_R$ charge to cancel gauge anomaly~\cite{Jung:2009jz, Ko:2013zsa, Nomura:2016emz,Nomura:2017tih,Chao:2017rwv} and forbid unnecessary Yukawa interactions to obtain inverse seesaw mechanism.
We then assign $A_4$ triplet representation to $S_L$ and $N_R$, and some relevant Yukawa couplings are written in terms of modular form 
providing a constrained structure of the neutrino mass matrix.
Then we perform numerical analysis scanning free parameters in the model and search for the region which can fit neutrino data.
For the allowed parameter sets, we show predictions in observables in the neutrino sector.

This paper is structured as follows. In Sec.\ref{sec:model} we briefly revisit the well-known inverse seesaw mechanism with discrete $A_4$-modular flavor symmetry and its appealing feature resulting in simple mass structures for charged leptons and neutral leptons including light active neutrinos and other two types of sterile neutrinos. We provide also the analytic formula for light neutrino masses and mixing along with the discussion on the non-unitarity effect. In Sec.\ref{sec:results} we study numerically the correlations between observables in the neutrino sector along with the input model parameters arising in $A_4$-modular symmetry and its predictions to lepton flavor violation. We briefly comment on  collider aspects of TeV scale pseudo-Dirac neutrinos in Sec.\ref{sec:collider} and conclude our results in Sec.\ref{sec:conclusion}.

\section{Model}
\label{sec:model}
We briefly discuss here the model framework for inverse seesaw mechanism introducing right-handed fermion under specific local Abelian symmetry $U(1)_R$ and modular $A_4$ symmetry. 
In the model, we introduce three families of right(left)-handed $SU(2)$ singlet fermions $N_R(S_L)$ with 1(0) charge under the $U(1)_R$ gauge symmetry,
and an isospin singlet boson $\varphi$ with 1 charge under the same $U(1)$ symmetry.
Furthermore, the SM Higgs boson $H$ also has charge 1 under $U(1)_R$ to induce the masses of SM fermions from the Yukawa Lagrangian after the spontaneous symmetry breaking. \footnote{Due to the feature of nonzero charges of $H$, lower bound on the breaking scale of $U(1)_R$ is determined via the precision test of $Z$ boson mass; $\Lambda \gtrsim {\cal O}$(10) TeV~\cite{Nomura:2017tih}.} \hspace*{-0.1cm}
Here we denote each of vacuum expectation value to be $\langle H\rangle\equiv v/\sqrt2$, and $\langle \varphi\rangle\equiv v'/\sqrt2$.
The scalar and gauge sector are the same as in Ref.~\cite{Nomura:2017tih} where $U(1)_R$ gauge boson mass is given by VEV of $\varphi$ and new gauge coupling $g_R$ as $m_{Z'} \simeq g_R v'$.
In this paper, we omit the details of the scalar sector and focus on the neutrino sector.

Using the particle content and symmetries mentioned in Table \ref{tab:fields-inverse},  
the relevant Yukawa Lagrangian for leptons--including charged leptons and neutral leptons-- can be written as, 
\begin{align}\label{a4lag}
 -\mathcal{L}_{\rm lepton} = \mathcal{L}_{M_\ell} + \mathcal{L}_{\rm 
M_D}+\mathcal{L}_{M}
+\mathcal{L}_{\rm \mu},
\end{align}
where $\mathcal{L}_{M_\ell} $ is interaction Lagrangian for charge leptons, $ \mathcal{L}_{\rm 
M_D}$ is for Dirac neutrino mass term connecting active light neutrinos $\nu_L$ and other sterile neutrino 
$N_R$, $ \mathcal{L}_{\rm M}$ is for mixing term between two types of sterile neutrinos $N_R$ and $S_L$ 
and $\mathcal{L}_{\rm \mu}$ is for Majorana mass term for sterile neutrino $S_L$. The Majorana mass terms for 
the sterile neutrinos $N_R$ and another term $\overline{L_L} H S^c_L $  connecting $\nu_L$ and $S_L$ are absent 
in the present framework due to appropriate $U(1)_R$ charge assignments. 
These Lagrangians should be invariant under $A_4$ symmetry and sum of modular weight should be zero for each term.

\begin{center} 
\begin{table}[tbh!]
\begin{tabular}{|c||c|c|c|c|c|c|c||c|c|c|}\hline\hline  
  & \multicolumn{7}{c||}{Fermions} & \multicolumn{2}{c|}{Bosons} \\ \hline \hline

& ~$Q_L$~& ~$u_R$~  & ~$d_R$~& ~$L_L$~& ~$e_R$~& ~$N_R$~& ~$S_L$~& ~$H$~& ~$\varphi$~ \\\hline\hline 
$SU(3)_C$ & $\bm{3}$  & $\bm{3}$ & $\bm{3}$ & $\bm{1}$ & $\bm{1}$ & $\bm{1}$ & $\bm{1}$ & $\bm{1}$ & $\bm{1}$   \\\hline 
$SU(2)_L$ & $\bm{2}$  & $\bm{1}$  & $\bm{1}$  & $\bm{2}$  & $\bm{1}$  & $\bm{1}$  & $\bm{1}$ & $\bm{2}$   & $\bm{1}$    \\\hline 
$U(1)_Y$   & $\frac16$ & $\frac23$ & $-\frac13$ & $-\frac12$  & $-1$ & $0$  & $0$  & $\frac12$ & $0$  \\\hline
$U(1)_{R}$   & $0$ & $1$ & $-1$   & $0$  & $-1$  & $1$  & $0$  & $1$ & $1$  \\\hline
$A_4$ & $\bm{1}$ & $\bm{1}$ & $\bm{1}$ & $\bm{1}, {\bf 1^{\prime \prime}}, {\bf 1^{\prime }}$ & ${\bf 1}, {\bf 1^{\prime }}, {\bf 1^{\prime \prime}}$ & $\bm{3}$ & $\bm{3}$ & $\bm{1}$ & $\bm{1}$ \\ \hline
$k_I$ & $0$ & $0$ & $0$ & $0$ & $0$ & $-2$ & $0$ & $0$ & $0$ \\
\hline
\end{tabular}
\caption{Particle content of the Standard Model extended with two types of sterile neutrinos $N_R, S_L$ and extra singlet scalar $\varphi$ for implementation of inverse seesaw mechanism and their charge assignments under $SU(2)_L\times U(1)_Y\times A_4$ where $k_I$ is the number of modular weight.}
\label{tab:fields-inverse}
\end{table}
\end{center}
\noindent \\
{\bf \underline{Dirac mass term for charged leptons ($M_{\ell}$):}}\\
In order to have a simplified structure for charged leptons mass matrix, we consider left-handed lepton doublets $\{L_{e_L}, L_{\mu_L}, L_{\tau_L} \}$ transforming under 
$A_4$ group as $\bm{1}, \bm{1}^{\prime \prime}, \bm{1}^{\prime}$ and right-handed charge leptons as $\bm{1}, \bm{1}^{ \prime}, \bm{1}^{\prime \prime}$ 
while SM Higgs doublets $H$ is transforming as singlet under $A_4$ group. All these fields are assigned zero modular weight and $0,1,1$ under $U(1)_R$ for 
left-handed lepton doublets, right-handed charged leptons and Higgs doublet, respectively. The relevant interaction Lagrangian term for charged leptons is given by
\begin{align}
 \mathcal{L}_{M_\ell}  
                   &= y_{\ell_{}}^{ee}  \overline{L}_{e_L} H e_R +  y_{\ell_{}}^{\mu \mu}  \overline{L}_{\mu_L} H \mu_R +  y_{\ell_{}}^{\tau \tau}  \overline{L}_{\tau_L} H \tau_R
                    + {\rm h.c.}, \label{Eq:yuk-Mell} 
\end{align}
After spontaneous symmetry breaking, the charged lepton mass matrix is found to be diagonal,
\begin{align}
M_\ell = \begin{pmatrix}  y_{\ell_{}}^{ee} v/\sqrt{2}  &  0 &  0 \\
                                       0  &  y_{\ell_{}}^{\mu \mu} v/\sqrt{2}  &  0 \\
                                       0  &  0  &  y_{\ell_{}}^{\tau \tau} v/\sqrt{2}        \end{pmatrix}  =
                     \begin{pmatrix}  m_e  &  0 &  0 \\
                                       0  &  m_\mu  &  0 \\
                                       0  &  0  &  m_\tau      \end{pmatrix}                   
\label{Eq:Mell} 
\end{align}

\noindent \\
{\bf \underline{Dirac neutrino mass term connecting $\nu_L$ and $N_R$ ($M_{D}$):}}\\
For flavor structure for Dirac neutrino mass matrix, we consider left-handed lepton doublets $\{L_{e_L}, L_{\mu_L}, L_{\tau_L} \}$ transforming under as $\bm{1}, \bm{1}^{\prime \prime}, \bm{1}^{ \prime}$ 
and right-handed neutrinos as triplet $\bm{3}$ under $A_4$ modular group while SM Higgs doublet $H$ is transforming as singlet under $A_4$ group. Then it is found that the generic Dirac Yukawa term 
$\overline{L_L} \tilde{H} N_R$ is protected by $A_4$ modular symmetry. The advantage of $A_4$ modular symmetry here is to allow this term without introducing additional fields while allowing the corresponding 
Yukawa coupling transforming under $A_4$ modular group as triplets shown in Table\,\ref{tab:couplings}. We use the modular forms of weight 2, {$\bm{Y}(\tau) = \left(y_{1}(\tau),y_{2}(\tau),y_{3}(\tau)\right)$},  transforming
as a triplet of $A_4$ which is given in terms of Dedekind eta-function  $\eta(\tau)$ and its derivative \cite{Feruglio:2017spp} which is given as Eq.~\eqref{eq:Y-A4} in the Appendix.

\begin{center} 
\begin{table}[t]
\begin{tabular}{|c||c|c|c|c|c|}\hline
{Yukawa coupling}  & ~{ $A_4$}~& ~$k_I$~     \\\hline 
{ $\bm{Y}$} & ${\bf 3}$ & ${\bf 2}$      \\\hline
\end{tabular}
\caption{Modular weight assignment for Yukawa coupling $\bm{Y}$ and its transformation under $A_4$ for giving flavor structure of different neutral fermion mass matrices.}
\label{tab:couplings}
\end{table}
\end{center}

As  a result of this, the relevant term for Dirac neutrino mass connecting light neutrinos $\nu_L$ and sterile neutrinos $N_R$ is given by
\begin{align}
 \mathcal{L}_{M_D}  
                   &= \alpha_D   \overline{L}_{e_L} \widetilde{H} (\bm{Y} N_R)_{1}   + \beta_D   \overline{L}_{\mu_L} \widetilde{H} (\bm{Y} N_R)_{1^{\prime}}
                   + \gamma_D   \overline{L}_{\tau_L} \widetilde{H} (\bm{Y} N_R)_{1^{\prime \prime}}                       + {\rm h.c.}, \label{Eq:yuk-MD} 
\end{align}
where subscript for the operator $\bm{Y} N_R$ indicates $A_4$ representation constructed by the product and $\{\alpha_D, \beta_D, \gamma_D\}$ are free parameters.
Using $\langle H \rangle = v/\sqrt{2}$,  the resulting Dirac neutrino mass matrix is found to be,
\begin{align}
M_D&=\frac{v}{\sqrt2}
\left[\begin{array}{ccc}
\alpha_D & 0 & 0 \\ 
0 & \beta_D & 0 \\ 
0 & 0 & \gamma_D \\ 
\end{array}\right]
\left[\begin{array}{ccc}
y_1 &y_3 &y_2 \\ 
y_2 &y_1 &y_3 \\ 
y_3 &y_2 &y_1 \\ 
\end{array}\right]_{LR}.                   
\label{Eq:Mell} 
\end{align}

\noindent \\
{\bf \underline{Mixing term connecting $N_R$ and $S_L$ ($M_{}$):}}\\
We chose both types of sterile neutrinos $N_R$ and $S_L$ transforming as  triplets $\bm{3}$ under $A_4$ modular group. However the mixing term 
$\overline{S_L} N_R$ is forbidden due to $U(1)_R$ charge assignment. Then this term is obtained from a Yukawa interaction with scalar singlet  $\varphi$ which has non-zero $U(1)_R$ charge and singlet under modular $A_4$ symmetry. The allowed mixing term for  $N_R$ and $S_L$ is given by
\begin{align}
 \mathcal{L}_{M}  
                   &= [\alpha_{SN} \bm{Y} (\overline{S_L} N_R)_{\rm symm} + \beta_{SN} \bm{Y} (\overline{S_L} N_R)_{\rm Anti-symm} ]\varphi^*   + {\rm h.c.} \nn \\
                   &=\alpha_{SN}[ y_1(2  \bar S_{L_1} N_{R_1} - \bar S_{L_2} N_{R_3} - \bar S_{L_3} N_{R_2})
+y_2(2  \bar S_{L_2} N_{R_2} - \bar S_{L_1} N_{R_3} - \bar S_{L_3} N_{R_1}) \nn \\
& \qquad \qquad + y_3(2  \bar S_{L_3} N_{R_3} - \bar S_{L_1} N_{R_2} - \bar S_{L_2} N_{R_1})] \varphi^* \nn\\
&+
\beta_{SN}[ y_1(  \bar S_{L_2} N_{R_3} - \bar S_{L_3} N_{R_2})
+y_2(  \bar S_{L_3} N_{R_1} - \bar S_{L_1} N_{R_3})+
y_3(  \bar S_{L_1} N_{R_2} - \bar S_{L_2} N_{R_1})] \varphi^* \nn \\ 
& + {\rm h.c.},                   \label{Eq:yuk-M} 
\end{align}
where first and second term in the first line correspond to symmetric and anti-symmetric product for $\bar S_L N_R$ making $\bm{3}$ representation of $A_4$.
Using $\langle \varphi \rangle = v'/\sqrt{2}$,  the resulting mass matrix is found to be,
\begin{align}
M_{NS}&=\frac{v'}{\sqrt2}
 \left(
 \alpha_{NS}\left[\begin{array}{ccc}
2y_1 & -y_3 & -y_2 \\ 
-y_3 & 2y_2 & -y_1 \\ 
-y_2 & -y_1 & 2y_3 \\ 
\end{array}\right]
+
\beta_{NS}
\left[\begin{array}{ccc}
0 &y_3 & -y_2 \\ 
-y_3 & 0 & y_1 \\ 
y_2 & -y_1 &0 \\ 
\end{array}\right]
\right).
\end{align}

\noindent \\
{\bf \underline{Majorana mass term for $S_L$ ($\mu_{}$):}}\\
Since the sterile neutrino $S_L$ transforming as  triplet $\bm{3}$ under $A_4$ modular group with zero modular weight, the Majorana mass term 
can be written as,
\begin{align}
 \mathcal{L}_{\mu}  
                   &= \mu_0 \overline{S^c_L} S_L               + {\rm h.c.}, \label{Eq:yuk-M} 
\end{align}
which results,
\begin{align}
\mu=  \mu_0 \begin{pmatrix}  1&  0 &  0 \\
                                       0  &  0 &  1 \\
                                      0  &  1 &  0       \end{pmatrix}  \, .              
\label{Eq:Mell} 
\end{align}

\subsection{Inverse Seesaw mechanism for light neutrino Masses}
Within the present model invoked with $A_4$ modular symmetry the complete $9 \times 9$ neutral fermion mass 
matrix for  inverse seesaw mechanism in the flavor basis of $\left(\nu_L, N^c_R, S_L \right)$ is given by
\begin{eqnarray}
\mathbb{M} = \left(\begin{array}{c|ccc}   & \nu_L & N^c_R  & S_L   \\ \hline
\nu_L  & 0       & M_D       & 0 \\
N^c_R    & M^T_D         & 0       & M^T_{NS} \\
S_L & 0     & M_{NS}    & \mu
\end{array}
\right).
\label{eq:numatrix-complete}
\end{eqnarray}

Using the appropriate mass hierarchy among mass matrices as given below~\footnote{The hierarchies among mass parameters could be explained by several mechanisms such as radiative models~\cite{Dev:2012sg, Dev:2012bd, Das:2017ski} and effective models with higher order terms \cite{Okada:2012np}.},
\begin{equation}
 M_{NS} > M_D >> \mu_{0} , 
\end{equation}
the inverse seesaw mass formula for light neutrinos is given by
\begin{eqnarray}
m_\nu 
&=&\left(\frac{M_D}{M_{NS}}\right) \mu \left(\frac{M_D}{M_{NS}}\right)^T\,. 
\end{eqnarray}
The above relation can be read as,
$$\left( \frac{m_\nu}{\mbox{0.1\, eV}}\right) = \left(\frac{M_D}{\mbox{100\, GeV}} \right)^2 
  \left(\frac{\mu}{\mbox{keV}}\right) \left(\frac{M_{NS}}{10^4\, \mbox{GeV}} \right)^{-2}\,.$$
Since mass parameters for $M_D,\mu,M_{NS}$ are overall factors,  we can define a dimensionless neutrino mass matrix $\tilde m_\nu$ as follows:
\begin{align}
m_\nu=\left(\frac{v}{v'}\right)^2\mu_0\tilde m_\nu\equiv \kappa\tilde m_\nu,\quad 
\kappa\equiv \left(\frac{v}{v'}\right)^2\mu_0. \label{eq:mu0}
\end{align}
Then, $\tilde m_\nu$ is diagonalized by $V^\dag_\nu (\tilde m_\nu^\dag \tilde m_\nu)V_\nu=(\tilde D_{\nu_1}^2,\tilde D_{\nu_2}^2,\tilde D_{\nu_3}^2)$. 
In this case, $\kappa$ is determined by
\begin{align}
(NO):\  \kappa^2= \frac{|\Delta m_{\rm atm}^2|}{\tilde D_{\nu_3}^2-\tilde D_{\nu_1}^2},
\quad
(IO):\  \kappa^2= \frac{|\Delta m_{\rm atm}^2|}{\tilde D_{\nu_2}^2-\tilde D_{\nu_3}^2},
 \end{align}
where $\Delta m_{\rm atm}^2$ is atmospheric neutrino mass difference squares, and NO and IO stand for normal and inverted ordering respectively.
Subsequently, the solar mass difference squares can be written in terms of $\kappa$ as follows:
\begin{align}
\Delta m_{\rm sol}^2= {\kappa^2}({\tilde D_{\nu_2}^2-\tilde D_{\nu_1}^2}),
 \end{align}
 which can be compared to the observed value.
For heavy sterile neutrino, we obtain pseudo Dirac mass for $\mu_0 \ll M_{NS}$ and mass eigenvalues are obtained by diagonalizing $M_{NS}$.
We write these eigenvalues as $M_{1,2,3}$ which will be numerically estimated.
  
In our model, one finds $U_{PMNS}=V_\nu$ since the charged-lepton is diagonal basis, and 
it is parametrized by three mixing angle $\theta_{ij} (i,j=1,2,3; i < j)$, one CP violating Dirac phase $\delta_{CP}$,
and two Majorana phases $\{\alpha_{21}, \alpha_{32}\}$ as follows:
\begin{equation}
U_{PMNS} = 
\begin{pmatrix} c_{12} c_{13} & s_{12} c_{13} & s_{13} e^{-i \delta_{CP}} \\ 
-s_{12} c_{23} - c_{12} s_{23} s_{13} e^{i \delta_{CP}} & c_{12} c_{23} - s_{12} s_{23} s_{13} e^{i \delta_{CP}} & s_{23} c_{13} \\
s_{12} s_{23} - c_{12} c_{23} s_{13} e^{i \delta_{CP}} & -c_{12} s_{23} - s_{12} c_{23} s_{13} e^{i \delta_{CP}} & c_{23} c_{13} 
\end{pmatrix}
\begin{pmatrix} 1 & 0 & 0 \\ 0 & e^{i \frac{\alpha_{21}}{2}} & 0 \\ 0 & 0 & e^{i \frac{\alpha_{31}}{2}} \end{pmatrix},
\end{equation}
where $c_{ij}$ and $s_{ij}$ stands for $\cos \theta_{ij}$ and $\sin \theta_{ij}$ respectively. 
Then, each of mixing is given in terms of the component of $U_{PMNS}$ as follows:
\begin{align}
\sin^2\theta_{13}=|(U_{PMNS})_{13}|^2,\quad 
\sin^2\theta_{23}=\frac{|(U_{PMNS})_{23}|^2}{1-|(U_{PMNS})_{13}|^2},\quad 
\sin^2\theta_{12}=\frac{|(U_{PMNS})_{12}|^2}{1-|(U_{PMNS})_{13}|^2}.
\end{align}
Also we compute the Jarlskog invariant and $\delta_{CP}$ derived from PMNS matrix elements $U_{\alpha i}$:
\begin{equation}
J_{CP} = \text{Im} [U_{e1} U_{\mu 2} U_{e 2}^* U_{\mu 1}^*] = s_{23} c_{23} s_{12} c_{12} s_{13} c^2_{13} \sin \delta_{CP},
\end{equation}
and the Majorana phases are also estimated in terms of other invariants $I_1$ and $I_2$:
\begin{equation}
I_1 = \text{Im}[U^*_{e1} U_{e2}] = c_{12} s_{12} c_{13}^2 \sin \left( \frac{\alpha_{21}}{2} \right), \
I_2 = \text{Im}[U^*_{e1} U_{e3}] = c_{12} s_{13} c_{13} \sin \left( \frac{\alpha_{31}}{2}  - \delta_{CP} \right).
\end{equation}
In addition, the effective mass for the neutrinoless double beta decay is given by
\begin{align}
\langle m_{ee}\rangle=\kappa|\tilde D_{\nu_1} \cos^2\theta_{12} \cos^2\theta_{13}+\tilde D_{\nu_2} \sin^2\theta_{12} \cos^2\theta_{13}e^{i\alpha_{21}}+\tilde D_{\nu_3} \sin^2\theta_{13}e^{i(\alpha_{31}-2\delta_{CP})}|,
\end{align}
where its observed value could be measured by KamLAND-Zen in future~\cite{KamLAND-Zen:2016pfg}. 
We will adopt the neutrino experimental data at 3$\sigma$ interval~\cite{Esteban:2018azc} as follows:
\begin{align}
&{\rm NO}: \Delta m^2_{\rm atm}=[2.431, 2.622]\times 10^{-3}\ {\rm eV}^2,\
\Delta m^2_{\rm sol}=[6.79, 8.01]\times 10^{-5}\ {\rm eV}^2,\\
&\sin^2\theta_{13}=[0.02044, 0.02437],\ 
\sin^2\theta_{23}=[0.428, 0.624],\ 
\sin^2\theta_{12}=[0.275, 0.350],\nn\\
&{\rm IO}: \Delta m^2_{\rm atm}=[2.413, 2.606]\times 10^{-3}\ {\rm eV}^2,\
\Delta m^2_{\rm sol}=[6.79, 8.01]\times 10^{-5}\ {\rm eV}^2,\\
&\sin^2\theta_{13}=[0.02067, 0.02461],\ 
\sin^2\theta_{23}=[0.433, 0.623],\ 
\sin^2\theta_{12}=[0.275, 0.350].\nn
\end{align}
We apply these ranges in searching for allowed parameter space in our numerical analysis.

\subsection{Non-unitarity}
Here, let us briefly discuss non-unitarity matrix $U'_{MNS}$.
This is typically parametrized by the form 
\begin{align}
U'_{MNS}\equiv \left(1-\frac12 FF^\dag\right) U_{MNS},
\end{align}
where $F\equiv  (M^{T}_{NS})^{-1} M_D$ is a hermitian matrix, and $U'_{MNS}$ represents the deviation from the unitarity. 
The global constraints are found via several experimental results such as the SM $W$ boson mass $M_W$, the effective Weinberg angle $\theta_W$, several ratios of $Z$ boson fermionic decays, invisible decay of $Z$, electroweak universality, measured Cabbibo-Kobayashi-Maskawa, and lepton flavor violations~\cite{Fernandez-Martinez:2016lgt}.
The result is then given by~\cite{Agostinho:2017wfs}
\begin{align}
|FF^\dag|\le  
\left[\begin{array}{ccc} 
2.5\times 10^{-3} & 2.4\times 10^{-5}  & 2.7\times 10^{-3}  \\
2.4\times 10^{-5}  & 4.0\times 10^{-4}  & 1.2\times 10^{-3}  \\
2.7\times 10^{-3}  & 1.2\times 10^{-3}  & 5.6\times 10^{-3} \\
 \end{array}\right].
\end{align} 
In our case, $F\equiv  (M^{T}_{NS})^{-1} M_D=\frac{v}{v'}(\tilde M^{T}_{NS})^{-1} \tilde m_D\approx \frac{v}{v'}$ if free parameters in $\tilde M_{NS}$ and $\tilde m_D$ are taken to be the same order.
Therefore, Non-unitarity can be controlled by $v'$ which is expected to be large mass scale.

\begin{figure}[tb!]\begin{center}
\includegraphics[width=80mm]{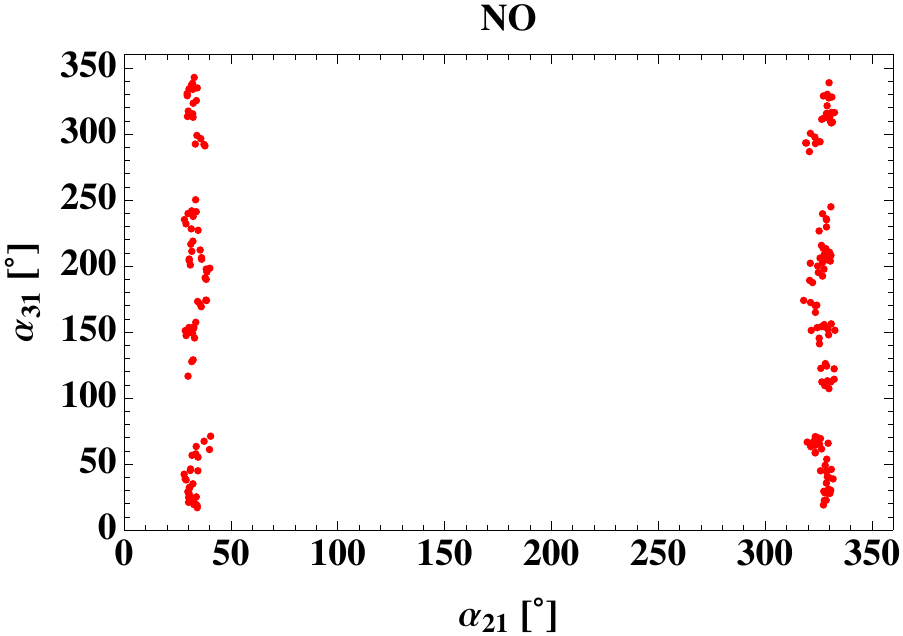}
\includegraphics[width=80mm]{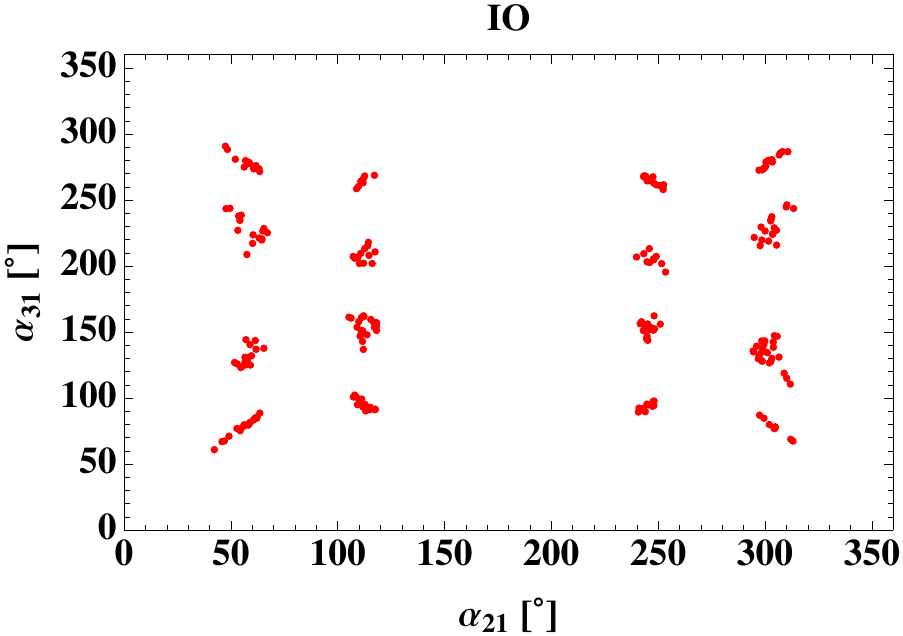}
\caption{The prediction of correlation between phases $\alpha_{21}$ and $\alpha_{31}$
where the left one is the case of NO and the right one is IO.}   
\label{fig:1}\end{center}\end{figure}
\section{Numerical analysis}
\label{sec:results}
In this section, we carry out numerical analysis.
Scanning free parameters in the model, we search for parameter sets satisfying neutrino data
and obtain some predictions in neutrino sector.

\begin{figure}[tb]\begin{center}
\includegraphics[width=80mm]{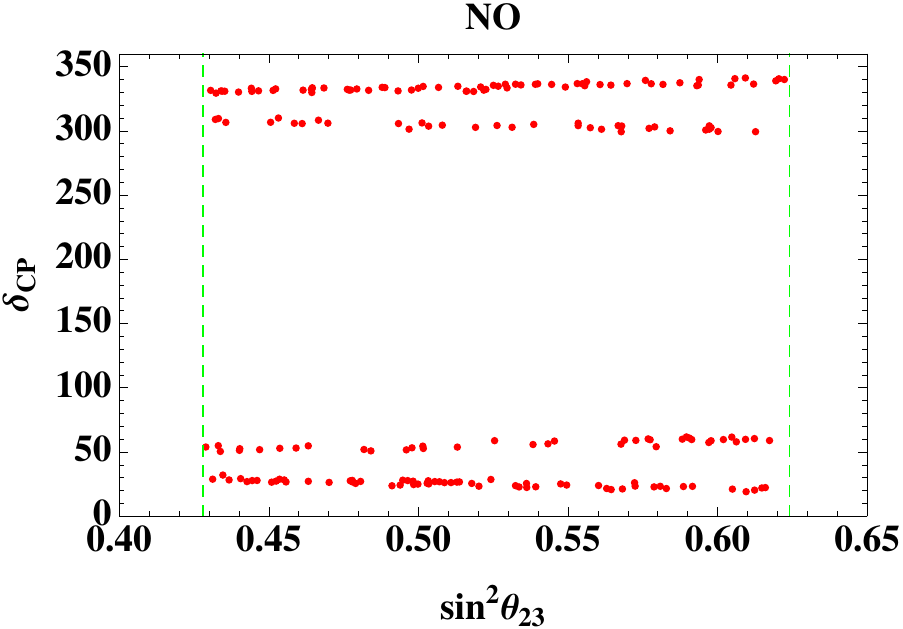} \
\includegraphics[width=80mm]{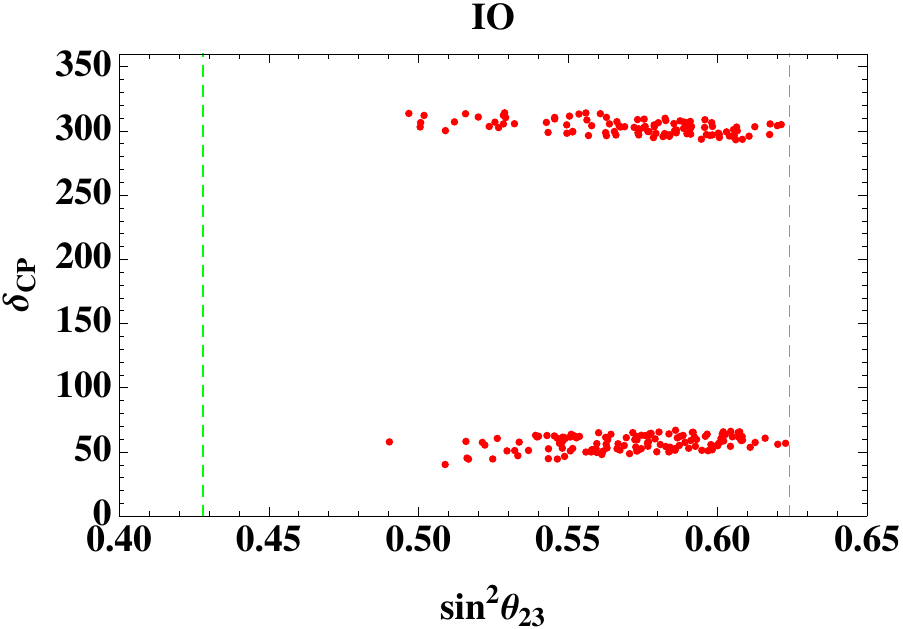} \\ \vspace{4mm}
\includegraphics[width=80mm]{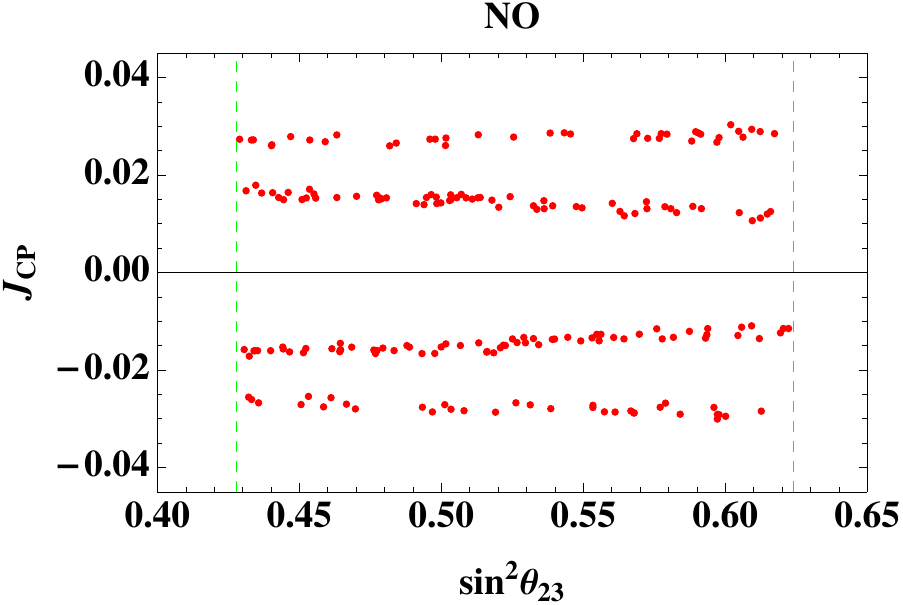} \
\includegraphics[width=80mm]{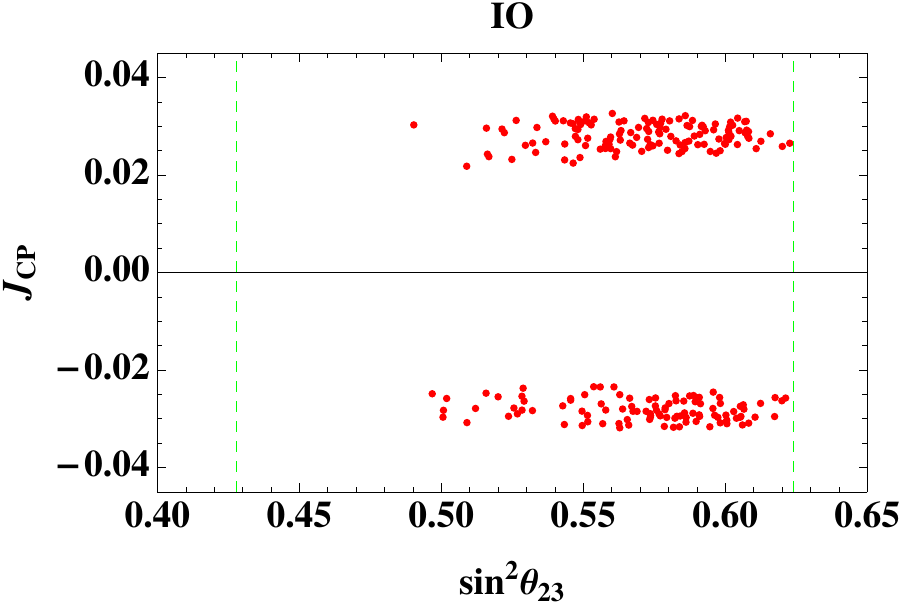}
\caption{Upper plots: the prediction of Dirac CP Phase $\delta^\ell_{CP}$ as a function of $\sin \theta_{23}$
where the left one is the case of NO and the right one is IO. 
Lower plots: The prediction of $J_{CP}$ as a function of $\sin \theta_{23}$
where the left one is the case of NO and the right one is IO}   
\label{fig:2}\end{center}\end{figure}

\noindent
{\bf \underline{Neutrino mass and mixing}} \\
Here we numerically analyze neutrino mass matrix applying the formulas in the previous section.
To fit neutrino data, we consider the free input parameters in following ranges:
\begin{align}
&|{\rm Re}[\tau]| \in [0,2],\quad {\rm Im}[\tau]\in [0.5,2],\quad \{ \alpha_{D,NS},\beta_{D,NS},\gamma_D \} \in [0.1,1.0],\nn\\
& v' \in [1.0,100] \ {\rm TeV},
\end{align}
where parameter $\mu_0$ is determined by Eq.~\eqref{eq:mu0} and not a free parameter.
Under these regions, we randomly scan the parameters and search for the allowed parameter sets satisfying all the neutrino oscillation data.

\begin{figure}[h!]\begin{center}
\includegraphics[width=80mm]{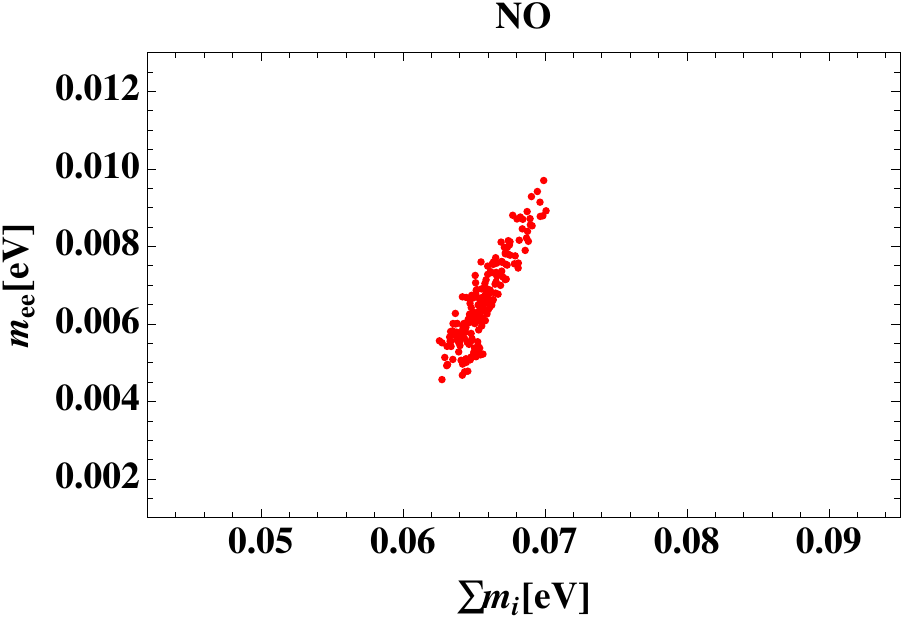}
\includegraphics[width=80mm]{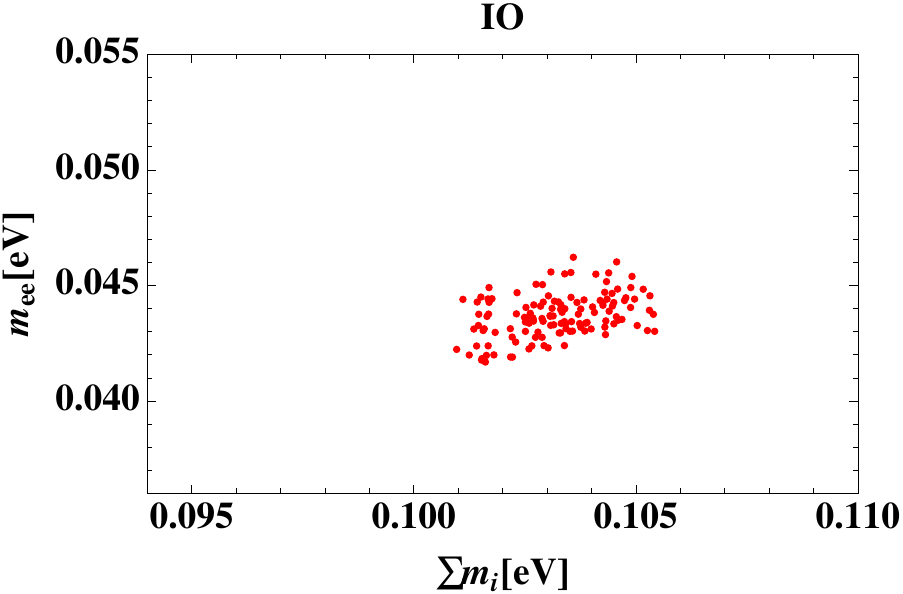}
\caption{The sum of neutrino masses $\sum m(\equiv \kappa {\rm Tr}[\tilde D_\nu]$) versus the effective mass for the neutrinoless double beta decay $\langle m_{ee}\rangle$,
where the left-handed figure is NO and the right one is IO.}   
\label{fig:3}\end{center}\end{figure}

As a result, we find parameter sets which can fit the neutrino data for both NO and IO cases.
The typical region of modulus $\tau$ is found in narrow space as -1.30 \ $\lesssim\ $Re$[\tau]\lesssim$\ -1.35 and  1.13 \ $\lesssim\ $Im$[\tau]\lesssim$\ 1.15 for NO 
and 0.99 \ $\lesssim\ | $Re$[\tau] |  \lesssim$\ 1.01 and  1.42 \ $\lesssim\ $Im$[\tau]\lesssim$\ 1.44 for IO.
Also ranges of $\mu_0$ are estimated as $\mu_0 \in [0.00001, 10]$ KeV for NO and $\mu_0 \in [0.01, 40]$ KeV.
Then, using the allowed parameter sets, we can obtain predictions such as for Dirac and Majorana phases.

Fig.~\ref{fig:1} shows correlation between two Majorana phases $\alpha_{21}$ and $\alpha_{31}$
where the left one is for the case of NO and the right one is for IO.
This figure implies that $\alpha_{21}$ favors $\sim 30^\circ$ or $\sim 330^\circ$ for NO,  
and $\sim 60^\circ$, $\sim 110^\circ$, $240^\circ$ or $300^\circ$  for IO.
On the other hand $\alpha_{31}$ can take values in wider range.

In Fig.~\ref{fig:2}, we find prediction on planes of $\{\sin^2 \theta_{23}, \delta_{CP} \}$ and $\{\sin^2 \theta_{23}, J_{CP} \}$ for both NO and IO cases.
We obtain the Dirac CP phases to be $\sim 30^\circ$, $\sim 50^\circ$ ,$\sim 310^\circ$ or $\sim 330^\circ$ for NO, and $\sim 50^\circ$ or $\sim 300^\circ$ for IO.
In addition, we predict $J_{CP} \sim \pm 0.015$ or $\pm 0.03$ for NO, and $J_{CP} \sim \pm (0.02-0.03)$ for IO.

\begin{figure}[h!]\begin{center}
\includegraphics[width=80mm]{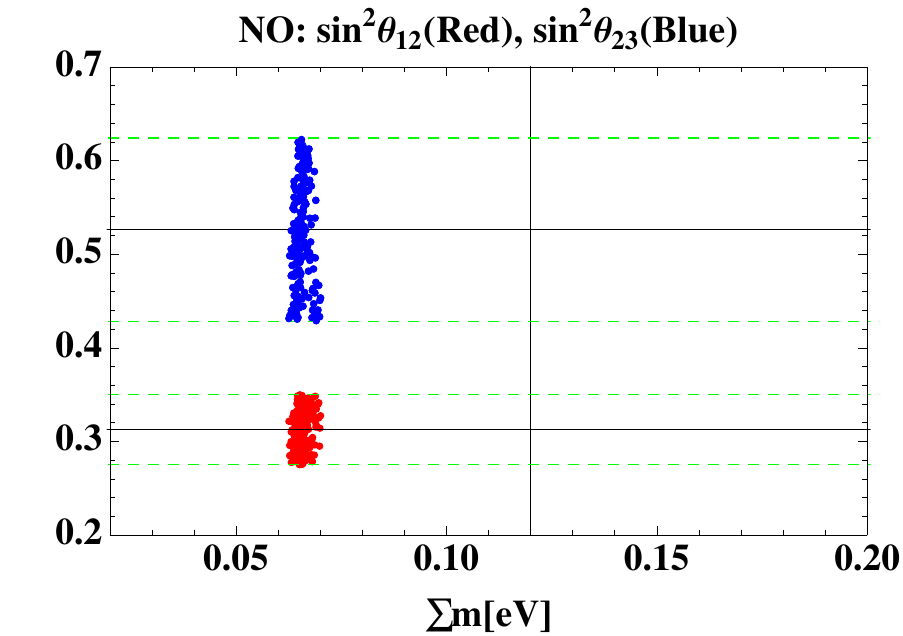}
\includegraphics[width=80mm]{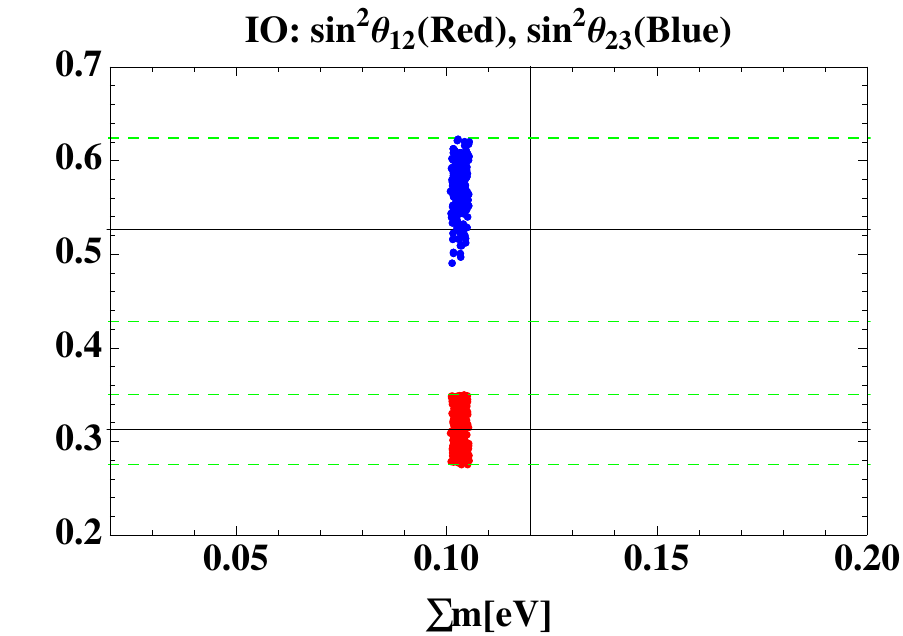}\\
\includegraphics[width=80mm]{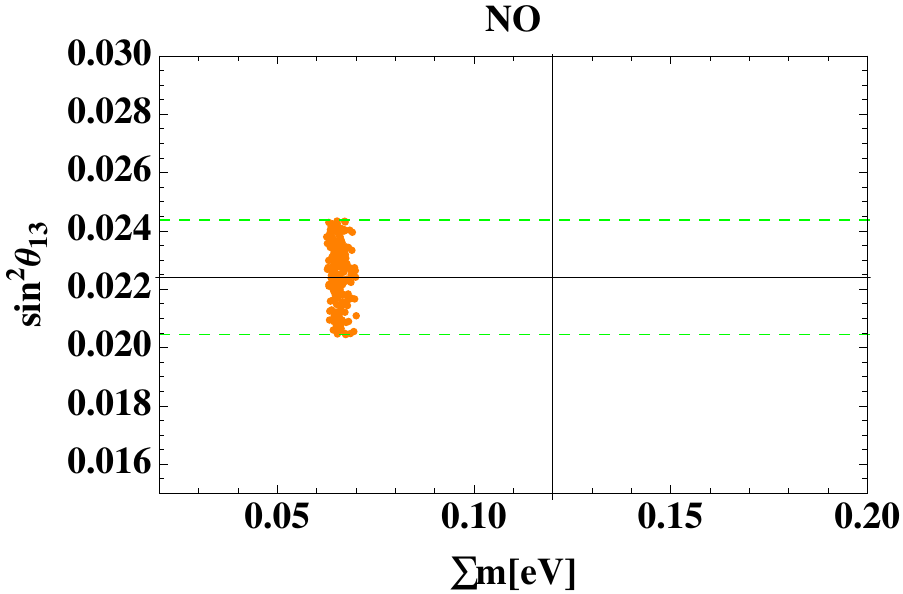}
\includegraphics[width=80mm]{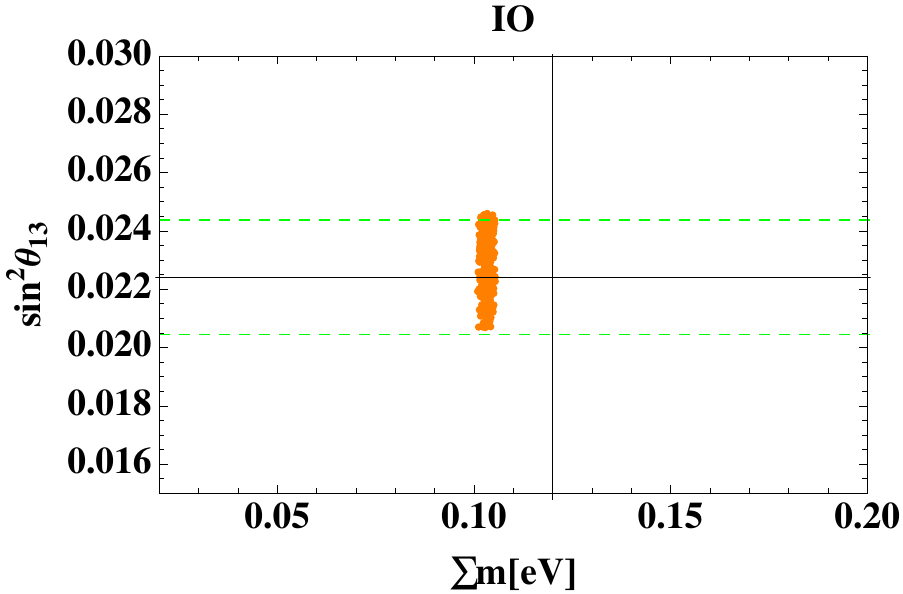}
\caption{The sum of neutrino masses $\sum m$ versus $\sin^2\theta_{12}(red), \sin^2\theta_{23}(blue)$ for top figures and $\sin^2\theta_{13}$ for bottom figures,
where the left-handed figure is NO and the right one is IO.}   
\label{fig:4}\end{center}\end{figure}

Fig.~\ref{fig:3} shows correlation between the sum of neutrino masses $\sum m(\equiv \kappa {\rm Tr}[\tilde D_\nu]$) versus the effective mass for the neutrinoless double beta decay $\langle m_{ee}\rangle$,
where the left-handed figure is for NO and the right one is for IO.
In case of NO, we have $0.06{\rm eV} \lesssim \sum m\lesssim0.07$eV and $0.0045{\rm eV} \lesssim \langle m_{ee}\rangle\lesssim0.01$eV.
In case of IO, we have $0.101{\rm eV} \lesssim \sum m\lesssim0.106$eV and $0.042{\rm eV} \lesssim \langle m_{ee}\rangle\lesssim0.046$eV.

Fig.~\ref{fig:4} shows relations between the sum of neutrino masses $\sum m(\equiv$ Tr$[ D_\nu]$) and $\sin^2\theta_{12} [\sin^2\theta_{23}]$ shown as red[blue] points for top figures, 
and $\sin^2\theta_{13}$ for bottom figures,
where the left-handed figure is for NO and the right one is for IO.
The allowed region of $\sin^2\theta_{23}$ in case of IO favors the second octant region [0.5,0.623],
which could be more precisely measured by the future experiment, although the other allowed regions run whole the experimental ranges.

\begin{figure}[tb]\begin{center}
\includegraphics[width=80mm]{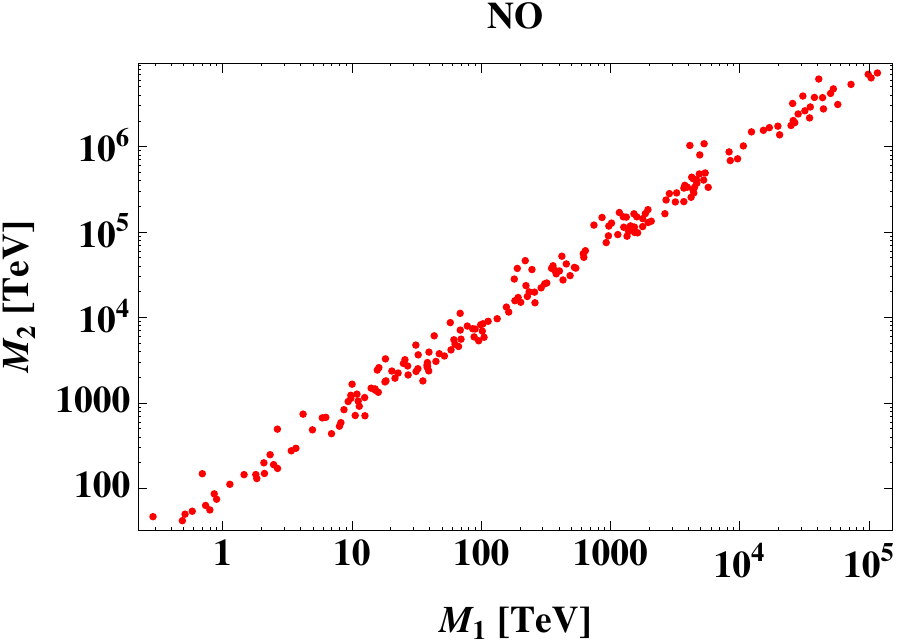}
\includegraphics[width=80mm]{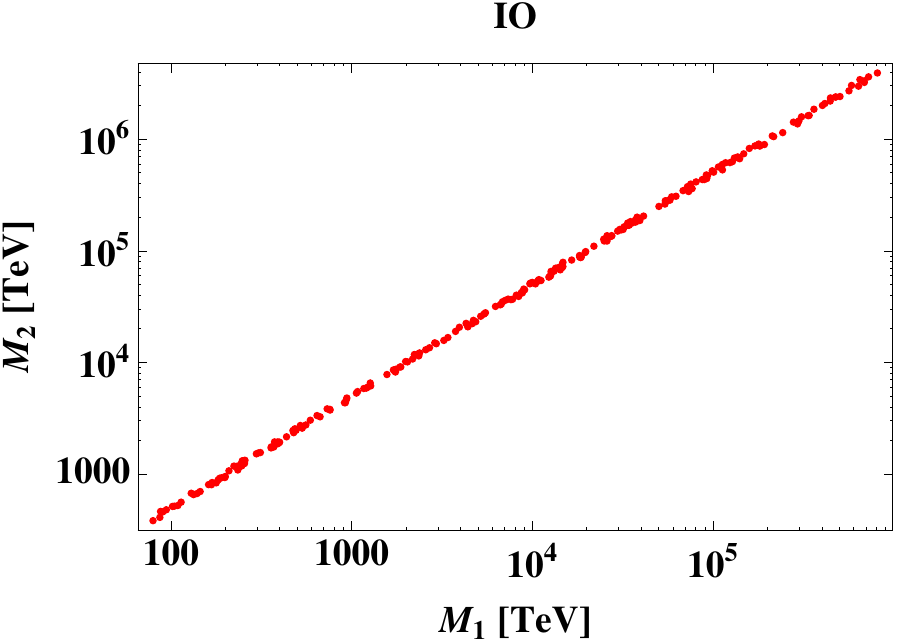}\\
\includegraphics[width=80mm]{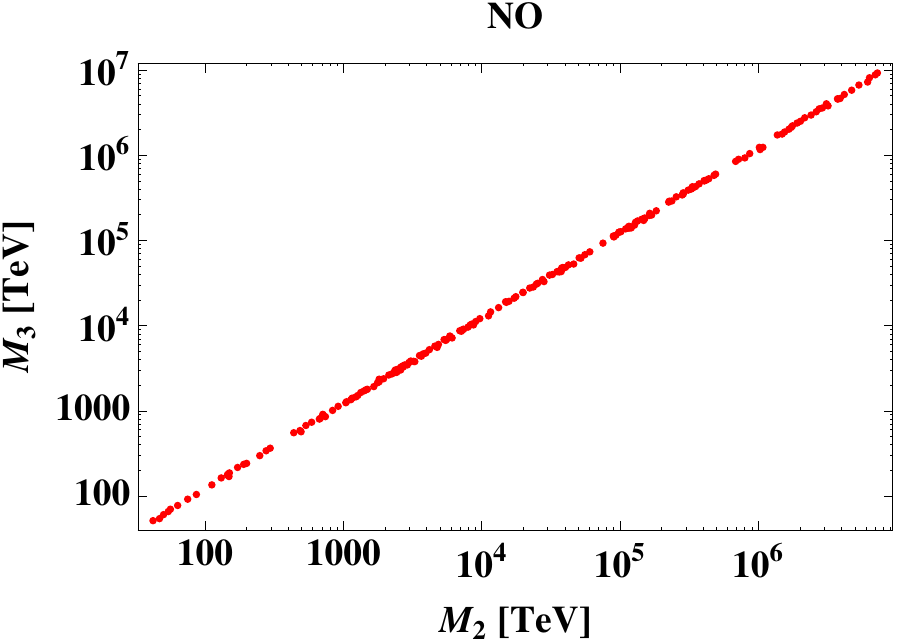}
\includegraphics[width=80mm]{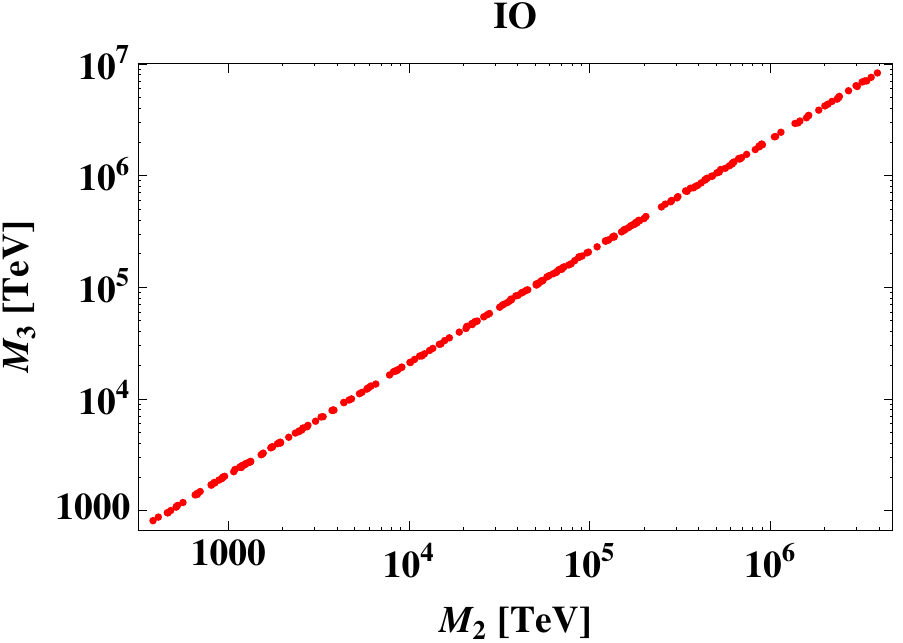}
\caption{Heavy neutrino masses where the left-handed figure is NO and the right one is IO.}   
\label{fig:5}\end{center}\end{figure}

In Fig.~\ref{fig:5}, we also show heavy neutrino masses where they are obtained as pseudo Dirac fermion. 
The mass relations can found as $M_1 \ll M_2 \sim M_3$ for NO and $M_1 < M_2 \lesssim M_3$ for IO.
In addition mass scale tends to be larger in IO case.

\noindent \\
{\bf \underline{Implication to lepton flavor violation}:-}
Experimental discovery of neutrino oscillation confirmed that neutrinos have non-zero masses and they mix. These observation also revealed that lepton flavor violation could possible in other low energy as well high energy experiments. Lepton flavor violating decays like $\mu\to e\gamma$, $\mu\to eee$ and $\mu\to e$~conversion in nuclei which are suppressed in SM by the GIM mechanism may have sizable contribution in the present model with sub-TeV pseudo-Dirac neutrinos. With large light-heavy neutrino mixing allowed in the present $A_4$-modular inverse seesaw mechanism and heavy pseudo-Dirac neutrino around few hundreds GeV mass range, the branching ratio for popular lepton flavor violating decay $\mu \to e \gamma$ is given by~\cite{Antusch:2014woa}
\begin{align}
	{\rm BR}(\mu \to e \gamma) = 
		\frac{3\alpha}{32\pi} \sum_{i=1}^3 f\left(\frac{M_i}{M_W}\right) 
		\left| F_{\mu i}^\ast \, F _{e i}\right|^2 \,.
\end{align}
Here, $M_i$ represents the heavy pseudo-Dirac neutrinos and $f(M^2_i/M^2_W)$ is a loop-function of order one. Also  $F\equiv  (M^{T}_{NS})^{-1} M_D=\frac{v}{v'}(\tilde M^{T}_{NS})^{-1} \tilde m_D \approx \frac{v}{v'}$ as discussed earlier when free parameters in $M_{D}$ and $M_{NS}$ are same order. 

With $M_i = 1$~TeV, the branching ratio for lepton flavor violating decay $\mu \to e \gamma$ is recasted in following way~\cite{Ibarra:2011xn}
\begin{align}
	{\rm BR}(\mu \to e \gamma) \simeq 
		8.4 \times 10^{-14} \cdot
		\left(\frac{|(F F^\dagger)_{e\mu}|}{10^{-5}}\right)^2\,.
\end{align}
The current bound derived from MEG experiment is ${\rm BR}(\mu \to e \gamma) < 4.2 \times 10^{-13}$ \cite{Adam:2013mnn,TheMEG:2016wtm}. This puts a limit on the ratio between the two VEVs, i.e $v/v^\prime \lesssim 10^{-2}$ when the sizes of $\{\alpha_{D,NS}, \beta_{D,NS}, \gamma_D \}$ are similar and $F \sim v/v'$. 
In that case, the translated bound on other singlet scalar field VEV is large as $v^\prime \gtrsim 10$ TeV
since $v$ is the usual SM Higgs VEV.  
We can relax the scale of $v^\prime$ by assuming hierarchy of free parameters in $M_{D}$ and $M_{NS}$, i.e. $\{\alpha_D, \beta_D, \gamma_D \} \ll \{ \alpha_{NS}, \beta_{NS} \}$. 
More parameter space in the model can be tested in future experiments.
There are other projected reach of sensitivity for LFV processes listed in Table\,\ref{tab:lfv-expt-bound} for
$\text{Br}\left(\tau \rightarrow e  \gamma \right), ~ \text{Br}\left(\tau \rightarrow 
\mu  \gamma \right), ~\text{Br}\left(\mu \rightarrow e 
 \gamma \right)$.  
\begin{table}[h!]
\centering
\begin{tabular}{|c|c|c|c|}
\hline \hline
 Branching ratio for LFV Decays              & Present expt. bound             & Future planned expt. sensitivity \\ 
\hline \hline
$\mbox{Br} \left(\mu \to {e\gamma} \right)$                 &$4.2\times 10^{-13}$ \cite{Adam:2013mnn,TheMEG:2016wtm}         &$6\times 10^{-14}$ \cite{Baldini:2013ke} \\[2mm]
\hline
$\mbox{Br} \left(\tau \to {e\gamma} \right)$   & $3.3\times 10^{-8}$  \cite{Aubert:2009ag}       &$3\times10^{-9}$ \cite{Aushev:2010bq}\\[2mm]
\hline
$\mbox{Br} \left(\tau \to {\mu\gamma} \right)$   & $4.4   \times 10^{-8}$ \cite{Aubert:2009ag}      & $3 \times 10^{-9}$ \cite{Aushev:2010bq} \\[2mm]
\hline
\end{tabular}
\caption{This table presents branching ratios for various lepton flavor violating processes, $\mu \to e \gamma$, $\tau \to e \gamma$ and $\tau \to \mu \gamma$ with their present experimental limit and projected future sensitivity. }
\label{tab:lfv-expt-bound}
\end{table}

\noindent
\section{Implication to collider physics}
\label{sec:collider}
We briefly comment here, without any numerical estimation, on most promising collider signature of heavy pseudo-Dirac neutrinos within present $A_4$-modular inverse seesaw mechanism from heavy Majorana neutrinos which can be feasible at LHC. The important process involving heavy pseudo-Dirac neutrinos which can be probed at collider is the trilepton plus missing energy as follows,
\begin{align}
\label{eq:A4modular-ISS}
	\sigma \left(pp \to N \ell^\pm \to \ell^\mp \ell^\pm +/\hspace*{-0.3cm}E \right) 
	&= 
	\sigma (pp \to W \to N \ell^\pm) 
	\times \mbox{Br} (N \to \ell^\mp \ell^\pm +
	/\hspace*{-0.3cm}E)
	\, . 
\end{align}

\begin{figure}[h!]
\centering
	\includegraphics[width=0.89\textwidth]{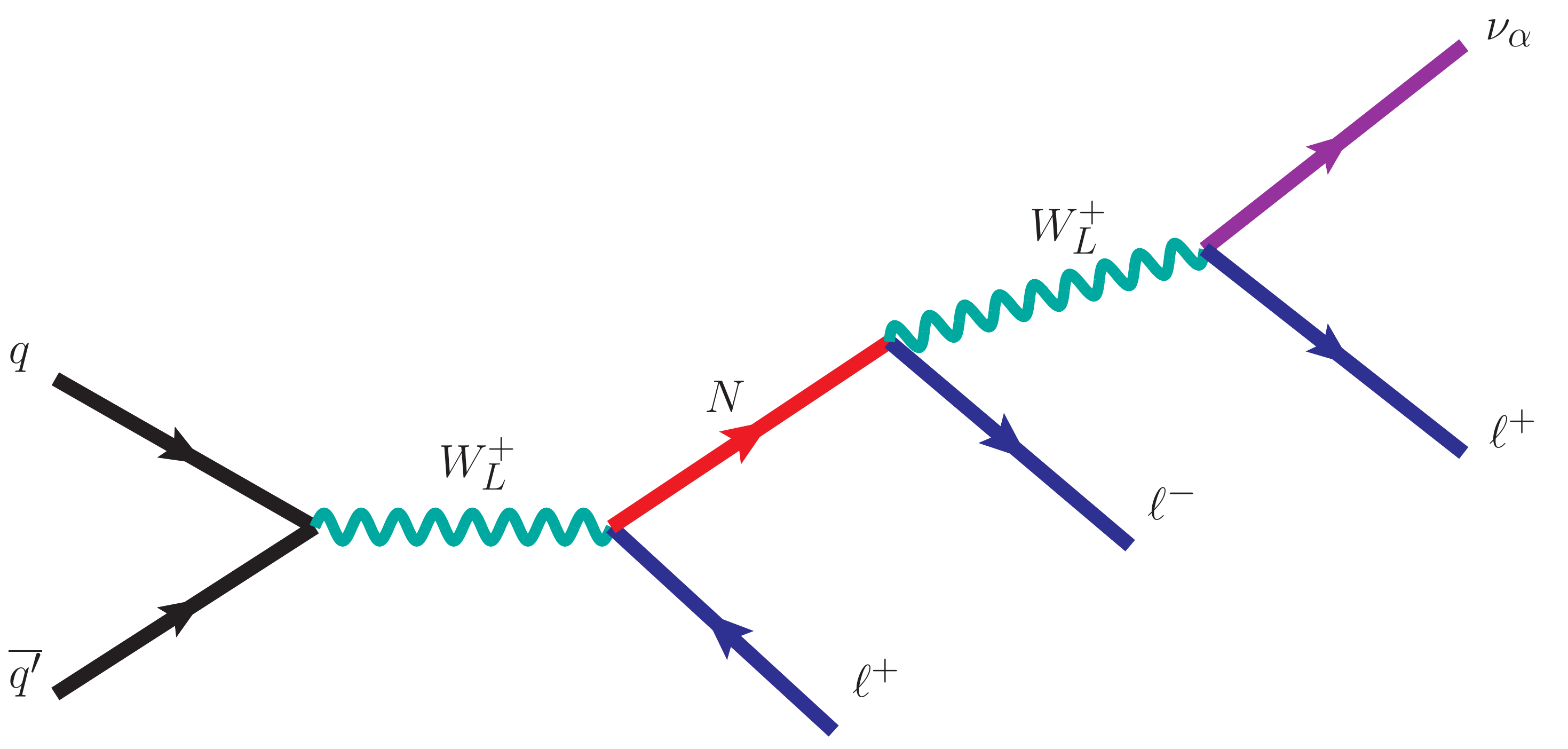}
\caption{Production of heavy pseudo-Dirac neutrinos arising in the presence $A_4$-modular inverse seesaw mechanism leading to trilepton plus missing energy signatures at Colliders. The Feynman diagram shows the $pp\to \ell^+ \ell^- \ell^+ /\hspace*{-0.3cm}E)$ process.}
\label{fig:samesign-dilepton}
\end{figure}
The feasibility of this trilepton plus missing energy at collider primarily depends on
\begin{itemize}
 \item large mixing between light active neutrinos and heavy pseudo-Dirac neutrinos,
 \item mass of the heavy pseudo-Dirac neutrinos, preferably, at few hundred GeV,
 \item production mechanism of this process.
\end{itemize}
The only way to distinguish between pseudo-Dirac from Majorana neutrinos at collider through careful analysis of their decay channels. In case of heavy Majorana neutrinos at TeV scale, like in type-I seesaw mechanism, the typical mixing between light-heavy neutrinos is $\theta_{\nu N}\simeq \sqrt{m_\nu M^{-1}_N}\leq 10^{-6}$ (see ref\,\cite{delAguila:2008cj} and references therein). In case of inverse seesaw mechanism with the introduction of small lepton number violating term $\mu$, the seesaw scale can be naturally in the testable range leading to large light-heavy neutrino mixing. At first, we produce heavy neutrinos, if kinematically allowed, through $q \overline{q^\prime} \to W^+_L \to \ell^+ N$ for heavy pseudo-Dirac neutrinos. After that heavy pseudo-Dirac neutrino decays to $\ell^- \ell^+ \nu_\alpha$ which crucially depends on large light-heavy neutrino mixing. 

\section{Conclusion}
\label{sec:conclusion}
We have studied an inverse seesaw model based on right-handed specific $U(1)$ gauge symmetry and modular $A_4$ symmetry where these symmetries forbid unnecessary terms and restrict structures of relevant Yukawa interactions. Majorana neutrino mass matrix has been formulated and it is characterized by modulus $\tau$ and some free parameters.

We have then carried out a numerical analysis to search for parameter sets that can fit neutrino oscillation data in both normal and inverted ordering cases.
For the allowed parameter sets, we find predictions such that; the Dirac CP phases to be $\sim 30^\circ$, $\sim 50^\circ$ ,$\sim 310^\circ$ or $\sim 330^\circ$ for NO, and $\sim 50^\circ$ or $\sim 300^\circ$ for IO;
sum of neutrino mass and $m_{ee}$ to be $0.06 (0.101)~{\rm eV} \lesssim \sum m\lesssim0.07(0.106)$~eV and $0.0045 (0.042)~{\rm eV} \lesssim \langle m_{ee}\rangle\lesssim0.01 (0.046)$~eV for NO(IO). 
In addition, we have shown hierarchy of heavy neutrino mass to be $M_1 \ll M_2 \sim M_3$ for NO and $M_1 < M_2 \lesssim M_3$ for IO.
Furthermore, we have discussed the implications of our model in lepton flavor violation and collider physics where we could test the model in future experiments.

\section*{Acknowledgments}
\vspace{0.5cm}
{\it
This research was supported by an appointment to the JRG Program at the APCTP through the Science and Technology Promotion Fund and Lottery Fund of the Korean Government. This was also supported by the Korean Local Governments - Gyeongsangbuk-do Province and Pohang City (H.O.). H. O. is sincerely grateful for the KIAS member.}

\section*{Appendix}

The modular group $\bar\Gamma$ is the group of linear fractional transformation
$\gamma$ acting on the modulus  $\tau$ 
belonging to the upper-half complex plane and its transformation is given as
\begin{equation}\label{eq:tau-SL2Z}
\tau \longrightarrow \gamma\tau= \frac{a\tau + b}{c \tau + d}\ ,~~
{\rm where}~~ a,b,c,d \in \mathbb{Z}~~ {\rm and }~~ ad-bc=1, 
~~ {\rm Im} [\tau]>0 ~ ,
\end{equation}
where it is isomorphic to  $PSL(2,\mathbb{Z})=SL(2,\mathbb{Z})/\{I,-I\}$ transformation.
This modular transformation is generated by two transformations $S$ and $T$ defined by 
\begin{eqnarray}
S:\tau \longrightarrow -\frac{1}{\tau}\ , \qquad\qquad
T:\tau \longrightarrow \tau + 1\ ,
\end{eqnarray}
and they satisfy the following algebraic relations, 
\begin{equation}
S^2 =\mathbb{I}\ , \qquad (ST)^3 =\mathbb{I}\ .
\end{equation}

Here we introduce the series of groups $\Gamma(N)~ (N=1,2,3,\dots)$ defined by
 \begin{align}
 \begin{aligned}
 \Gamma(N)= \left \{ 
 \begin{pmatrix}
 a & b  \\
 c & d  
 \end{pmatrix} \in SL(2,\mathbb{Z})~ ,
 ~~
 \begin{pmatrix}
  a & b  \\
 c & d  
 \end{pmatrix} =
  \begin{pmatrix}
  1 & 0  \\
  0 & 1  
  \end{pmatrix} ~~({\rm mod} N) \right \}
 \end{aligned} .
 \end{align}
Here we define $\bar\Gamma(2)\equiv \Gamma(2)/\{I,-I\}$ for $N=2$.
Since the element $-I$ does not belong to $\Gamma(N)$
  for $N>2$ case, we have $\bar\Gamma(N)= \Gamma(N)$,
  which are infinite normal subgroup of $\bar \Gamma$ known as principal congruence subgroups.
   Then finite modular groups are obtained as the quotient groups defined as
   $\Gamma_N\equiv \bar \Gamma/\bar \Gamma(N)$.
For these finite groups $\Gamma_N$, $T^N=\mathbb{I}$  is imposed.
 Then the groups $\Gamma_N$ with $N=2,3,4,5$ are isomorphic to
$S_3$, $A_4$, $S_4$ and $A_5$, respectively \cite{deAdelhartToorop:2011re}.

Modular forms of level $N$ are 
holomorphic functions $f(\tau)$ which are transformed under the action of $\Gamma(N)$ as follows:
\begin{equation}
f(\gamma\tau)= (c\tau+d)^k f(\tau)~, ~~ \gamma \in \Gamma(N)~ ,
\end{equation}
where $k$ is the so-called as the  modular weight.

Here we discuss the modular symmetric theory without imposing supersymmetry explicitly. 
In this paper, we consider the $A_4$ ($N=3$) modular group. 
Under the modular transformation of Eq.(\ref{eq:tau-SL2Z}), a field $\phi^{(I)}$ 
transforms such that 
\begin{equation}
\phi^{(I)} \to (c\tau+d)^{-k_I}\rho^{(I)}(\gamma)\phi^{(I)},
\end{equation}
where  $-k_I$ is the modular weight and $\rho^{(I)}(\gamma)$ denotes an unitary representation matrix of $\gamma\in\Gamma(2)$.

The kinetic terms of scalar fields can be written by 
\begin{equation}
\sum_I\frac{|\partial_\mu\phi^{(I)}|^2}{(-i\tau+i\bar{\tau})^{k_I}} ~,
\label{kinetic}
\end{equation}
which is invariant under the modular transformation and overall factor is eventually absorbed by a field redefinition.
Then the Lagrangian should be invariant under the modular symmetry.

The modular forms with weight 2, {\bf Y} = {$(y_{1},y_{2},y_{3})$},  transforming
as a triplet of $A_4$ is written in terms of Dedekind eta-function  $\eta(\tau)$ and its derivative \cite{Feruglio:2017spp}:
\begin{eqnarray} 
\label{eq:Y-A4}
y_{1}(\tau) &=& \frac{i}{2\pi}\left( \frac{\eta'(\tau/3)}{\eta(\tau/3)}  +\frac{\eta'((\tau +1)/3)}{\eta((\tau+1)/3)}  
+\frac{\eta'((\tau +2)/3)}{\eta((\tau+2)/3)} - \frac{27\eta'(3\tau)}{\eta(3\tau)}  \right), \nonumber \\
y_{2}(\tau) &=& \frac{-i}{\pi}\left( \frac{\eta'(\tau/3)}{\eta(\tau/3)}  +\omega^2\frac{\eta'((\tau +1)/3)}{\eta((\tau+1)/3)}  
+\omega \frac{\eta'((\tau +2)/3)}{\eta((\tau+2)/3)}  \right) , \label{eq:Yi} \\ 
y_{3}(\tau) &=& \frac{-i}{\pi}\left( \frac{\eta'(\tau/3)}{\eta(\tau/3)}  +\omega\frac{\eta'((\tau +1)/3)}{\eta((\tau+1)/3)}  
+\omega^2 \frac{\eta'((\tau +2)/3)}{\eta((\tau+2)/3)}  \right)\,.
\nonumber
\end{eqnarray}
%
 Notice here that any singlet couplings under $A_4$ start from $-k=4$ while they are zero if $-k=2$.


\end{document}